# Clarifying ultrafast carrier dynamics in ultrathin films of the topological insulator Bi$_2$Se$_3$ using transient absorption spectroscopy

Yuri D. Glinka [a,b,*], Junzi Li [c], Tingchao He [c,*], Xiao Wei Sun [a,d,*]

[a] Guangdong University Key Lab for Advanced Quantum Dot Displays and Lighting, Shenzhen Key Laboratory for Advanced Quantum Dot Displays and Lighting, Department of Electrical and Electronic Engineering, Southern University of Science and Technology, Shenzhen 518055, China.
[b] Institute of Physics, National Academy of Sciences of Ukraine, Kiev 03028, Ukraine.
[c] College of Physics and Energy, Shenzhen University, Shenzhen 518060, China.
[d] Shenzhen Planck Innovation Technologies Pte Ltd., Longgang, Shenzhen 518112, China

**Abstract**

Ultrafast carrier dynamics in the topological insulator Bi$_2$Se$_3$ have recently been intensively studied using a variety of techniques. However, we are not aware of any successful experiments exploiting transient absorption (TA) spectroscopy for these purposes. Here we demonstrate that if the ~730 nm wavelength pumping (~1.7 eV photon energy) is applied to ultrathin Bi$_2$Se$_3$ films, TA spectra cover the entire visible region, thus unambiguously pointing to two-photon excitation (~3.4 eV). The carrier relaxation dynamics is found to be governed by the polar optical phonon cascade emission occurring in both the bulk states and the Dirac surface states (SS), including SS-bulk-SS vertical electron transport and being also exclusively influenced by whether the Dirac point is presented between the Dirac cones of the higher energy (~1.5 eV) Dirac SS (known as SS2). We have recognized that SS2 act as a valve substantially slowing down the relaxation of electrons when the gap between Dirac cones exceeds the polar optical phonon and resonant defects energies. The resulting progressive accumulation of electrons in the gapped SS2 becomes detectable through the inverse bremsstrahlung type free carrier absorption.

*Corresponding authors
E-mail addresses: yuridglinka@yahoo.com (Y.D. Glinka), tche@szu.edu.cn (T. He), sunxw@sustech.edu.cn (X.W. Sun)

## 1. Introduction

The topologically protected gapless Dirac surface states (SS) in topological insulators (TI) are novel quantum states of matter that are being caused by a combination of strong spin-orbit interaction and time-reversal symmetry.[1,2] Dirac SS reveal linear dispersion resulting in the emergence of two Dirac cones joined through the Dirac point and governing the metallic-like surface conductivity of these materials. The TI $Bi_2Se_3$ is characterized by a pair of such Dirac SS energetically distanced by ~1.5 eV and known as SS1 and SS2.[3] Consequently, SS1 occur in the midgap of $Bi_2Se_3$ (bandgap $E_g$ ~ 0.3 eV), usually being partially occupied due to natural *n*-doping. In contrast, SS2 emerge between the higher energy unoccupied bulk states.[3] The situation becomes even more complicated when switching to ultrathin $Bi_2Se_3$ films of a few quintuple layers (QL) thickness (QL ~ 0.954 nm). Due to coupling between Dirac SS on the opposite surfaces of the film, the energy gap in the topologically protected SS1 and SS2 can be formed. Consequently, depending on the film thickness, one can recognize three topological phases: (i) the gapless 3D TI phase (6 – 10 QL), (ii) the gapped 2D hybrid TI (HTI) phase (3 – 5 QL), and (iii) the gapped 2D topologically trivial insulator (TTI) phase (1 – 2 QL).[4,5] Furthermore, quantum confinement additionally modifies the band structure of $Bi_2Se_3$ films thinner than 10 QL, progressively enhancing the effect of the quantum well discrete states with decreasing film thickness.[3,5] Owing to such a complicated band structure and the film thickness dependent topological phase transitions, the ultrafast relaxation of highly energetic carriers in ultrathin $Bi_2Se_3$ films through the longitudinal optical (LO) phonon cascade emission (Fröhlich interaction)[6,7] is also expected to include the SS-bulk-SS vertical carrier transport.[8] The resulting spatial variations of the carrier density are expected to affect the surface conductivity of the TI $Bi_2Se_3$ and therefore the deep understanding of this dynamics remains one of the most challenging problems governing potential applications of the TI $Bi_2Se_3$ in novel electronic and optoelectronic devices.

The ultrafast carrier dynamics in the TI $Bi_2Se_3$ have recently been intensively studied using a variety of techniques, such as time-resolved and angle-resolved photoemission spectroscopy (TrARPES),[6,9-12] pump-probe reflectivity,[13-18] pump-probe second harmonic generation (SHG),[8,14] transient reflectivity spectroscopy,[19-21] and transient conductivity spectroscopy.[22,23] In TrARPES experiments, the fundamental beam of a commercial ultrafast

Ti:Sapphire laser (~800 nm - 1.55 eV photon energy) is usually applied as a pumping beam, whereas the UV range probing beam allows photoemitting electrons in the one-photon regime as well.[6,9-12] Similarly, the identical wavelength pumping and probing beams of a Ti:Sapphire laser have been used in the most common pump-probe reflection and SHG configurations.[8,13-18] On the contrary, in transient reflectivity spectroscopy, the visible range pumping and either the visible or mid-IR/THz range probing were applied.[19-21] Finally, the visible/mid-IR/THz range pumping and the THz range probing have been used in transient conductivity spectroscopy.[22,23]

The TrARPES technique is known to image the transiently excited carrier population within the 2 – 3 nm length scale near the sample surface.[6,9-12] The pump-probe SHG technique is also surface sensitive, probing charge-separation dynamics within a few atomic layers through the electric-field-induced SHG response.[8,14,24] In contrast, the visible and mid-IR/THz range transient reflectivity and transient conductivity techniques monitor the transiently excited carrier population on the much longer length scale of a few tens of nanometers through the Pauli blocking mechanism (photoinduced bleaching) and through either the bandgap renormalization or free carrier absorption (FCA) mechanism (photoinduced absorption).[13-23,25] The FCA mechanism, however, is known to appear differently in the mid-IR range (Drude absorption) and in the visible range (inverse bremsstrahlung absorption).[25] These considerations also indicate that the mid-IR/THz range probing is usually applied to monitor carrier population dynamics directly in SS1, whereas the visible range probing is capable of monitoring carrier population dynamics directly in the higher energy bulk states and SS2. However, because the pumping beam induces the complex refractive index modulation in a wide spectral range energetically covering both SS1 and SS2, the visible range probing can also be applied to monitor ultrafast carrier dynamics in SS1.[8]

Specifically, if carriers are photoexcited in the TI $Bi_2Se_3$ films by photons with energies exceeding the Dirac point of SS2 (~1.5 eV)[3] using a commercial Ti:Sapphire laser, for example, (Fig. 1) the elucidated carrier relaxation trends include (i) the rapid electron-electron thermalization (~0.2 - 0.3 ps),[8,13-18] (ii) the film thickness dependent electron relaxation through the LO-phonon cascade emission (~1.5 - 3.5 ps)[16] (Fig. 1), (iii) the film thickness dependent metastable population of the upper Dirac cone of SS2 by relaxing electrons and their radiative/non-radiative recombination with holes residing in SS1 (~5.0 – 150 ps)[17] (Fig. 1), (iv) a metastable population of the conduction band (CB) edge (≤3.5 ps), continuously feeding the upper Dirac cone of SS1, and (v) the radiative/non-radiative recombination of electrons and holes residing in the upper and lower Dirac cones of SS1, respectively (a few nanoseconds).[8, 13-18] It is important to point out that because two-photon pumping has never been applied, the ultrafast dynamics of carrier relaxation from the bulk states and Dirac SS with energies higher than 1.55 eV has not been studied (Fig. 1).

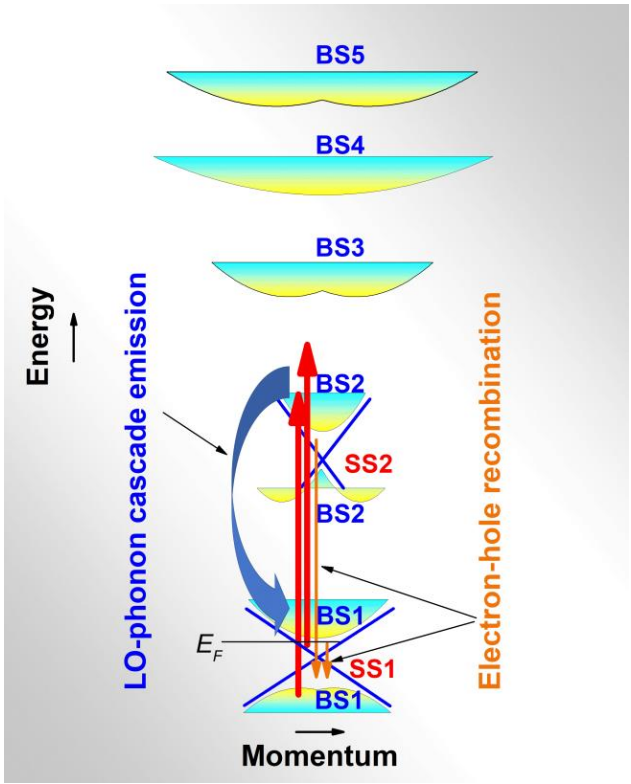

**Fig. 1.** The schematic band structure representation of 3D TI $Bi_2Se_3$ (~10 QL thick $Bi_2Se_3$ film), which includes the bulk states (BS1, BS2, BS3, BS4, and BS5) and Dirac surface states (SS1 and SS2). BS1 corresponds to the bulk valence band and the lowest energy bulk conduction band. The red color vertical up arrows demonstrate the one-photon carrier excitation with ~800 nm pumping (~1.55 eV photon energy) from the bulk valence band and from the upper Dirac cone of SS1 below the Fermi level ($E_F$). The blue color wide down arrow shows the relaxation of photoexcited electrons through the LO-phonon cascade emission, while the orange color vertical down arrows represent the electron-hole recombination between holes residing in the lower Dirac cone of SS1 and electrons residing in the upper Dirac cone of both SS2 and SS1. The higher energy bulk states (BS3, BS4, and BS5) remain unoccupied since the two-photon pumping regime has never been considered.

It is worth noting that no successful experiments exploiting transient absorption (TA) spectroscopy to study ultrafast carrier relaxation dynamics in the TI $Bi_2Se_3$ were reported. The reason seems to be that $Bi_2Se_3$ films are required to be high quality and thin enough to transmit broad range probing light without significant attenuation and distortion. The high quality and highly transparent in the broad spectral range thin substrates are also required for these purposes.

Here we repot on the application of TA spectroscopy to study ultrafast carrier dynamics in ultrathin $Bi_2Se_3$ films corresponding to different topological phases. Specifically, we studied three samples which were grown on thin (0.3 mm) sapphire substrates using molecular beam epitaxy and referred to as the 3D TI phase (10 QL), the 2D HTI phase (4 QL), and the 2D TTI phase (2 QL). We found that if the ~730 nm wavelength pumping (~1.7 eV photon energy) is applied, TA spectra cover the entire visible region, thus unambiguously pointing to two-photon excitation with a total energy of ~3.4 eV. Because the pumping photon energy is close to the Dirac point energy of SS2 (~1.5eV)[3], the pumping photons are suggested to resonantly excite electrons initially residing in the upper Dirac cone of SS1 below the Fermi level ($E_F$) toward the higher energy SS (Fig. 1). The relaxation of two-photon-excited electrons is found to occur initially through the LO-phonon cascade emission in the higher energy SS, followed by the LO-phonon-assisted SS-bulk-SS vertical electron transport, and finally being exclusively influenced by whether the Dirac point is presented between the Dirac cones of SS2. The Dirac SS2 act hence as a valve substantially slowing down the relaxation of electrons when the gap between the Dirac cones exceeds the LO-phonon and resonant defects energies. This kind of dynamics leads to the progressive accumulation of electrons in the upper Dirac cone of the gapped SS2 of the 2D HTI and 2D TTI phases. The latter process is evidenced to become detectable through the inverse bremsstrahlung type FCA. Finally, we concluded that TA spectroscopy is a powerful tool for unraveling carrier relaxation dynamics in ultrathin films of the TI $Bi_2Se_3$ of different topological phases.

## 2. Experimental Section

### 2.1. Sample preparation

The $Bi_2Se_3$ ultrathin films of 2 QL, 4 QL, and 10 QL thick (~2 nm, ~4 nm, and ~10 nm, respectively) were grown on the 0.3 mm $Al_2O_3$(0001) substrates using molecular beam epitaxy, with a 10 nm thick $MgF_2$ protecting capping layer, which was grown at room temperature without exposing the film to atmosphere. The quality of the films, their structure and the thickness estimation procedure were discussed elsewhere.[26] Specifically, the samples have been found to be epitaxial and the nominal number of QL was accurate to approximately 5%. The disorder has been found to increase with the sample thickness, but all samples less than or equal to 20 QL in thickness were of similar structural quality.

**2.1. Experimental setup**

TA spectra were measured using the Transient Absorption Spectrometer (Newport), which was equipped with a Spectra-Physics Solstice Ace regenerative amplifier (~800 nm wavelength, ~100 fs pulses with 1.0 KHz repetition rate) for the generation of the probing beam supercontinuum within the entire visible region (1.65 - 3.9 eV) and with the Topas light convertor for the pumping beam. The spectrometer was also additionally modified to suppress all the coherent artifacts originating from the sapphire substrate in the sub-picosecond timescale. We used optical pumping at 730 nm (1.7 eV photon energy). The probing supercontinuum beam was at normal incidence, whereas the pumping beam was at an incident angle of ~30°. All measurements were performed in air and at room temperature using a cross-linear-polarized geometry [the pumping and probing beams were polarized out-of-plane (vertical) and in-plane (horizontal) of incidence, respectively]. The data matrix was corrected for the chirp of the supercontinuum probing pulse and the zero-time was adjusted for each measurement using the coherent signal from a thin (0.3 mm) sapphire plate.

The spot sizes of the pumping and probing beams were ~400 μm and ~150 μm, respectively. The pumping beam average power has ranged from ~ 1.0 mW to ~10.0 mW (the corresponding pumping pulse power density has ranged from ~ 8.0 GW cm$^{-2}$ to ~80 GW cm$^{-2}$). The broadband probing beam was of the ~0.4 mW power, which for the similar to the pumping beam bandwidth (~26 meV) provides the probing pulse power density of ~0.15 GW cm$^{-2}$. Because the latter value is much smaller than that of the pumping pulse, the probing beam effect on carrier excitation is expected to be negligible.

**3. Results and discussion**

**3.1 TA spectra of ultrathin Bi$_2$Se$_3$ films of different topological phases: the electron density effect.**

Figure 2(a) and (c) shows the pseudo-color TA spectra plots of ultrathin Bi$_2$Se$_3$ films, which were induced by an ultrashort (~100 fs) ~730 nm pumping pulse (~1.7 eV photon energy). The TA spectra cover the entire visible region, thus unambiguously suggesting that two-photon pumping (~3.4 eV) occurs. According to the Franck–Condon principle[27] and because we used unintentionally *n*-doped Bi$_2$Se$_3$ films,[28] the two-photon pumping is expected to occur as a vertical transition that excites electrons initially residing in the upper Dirac cone of SS1 below $E_F$ toward the higher energy SS (Fig. 1). This kind of the surface-to-surface two-photon direct transition seems to be more probable compared to the surface-to-bulk two-photon direct transition for two reasons. First, the Dirac SS in ultrathin Bi$_2$Se$_3$ films is known to be the continuum type compared to the quantum well discrete type bulk states.[3,5] As a result, two-photon excitation can always occur through the SS2 resonance, whereas the bulk resonance should be tuned especially. Secondly, the two-photon absorption requires a short coherence time between two photons to be absorbed sequentially.[29] This coherence time is governed by a shift of the equilibrium lattice coordinates, which is induced by an electron photoexcited by the first photon in an intermediate state, by an amount proportional to the electron–phonon coupling strength. Consequently, the probability of two-photon excitation in Dirac SS is high enough due to exceptionally weak electron-phonon coupling.[30] Alternatively, because electron-phonon coupling in the bulk states is much stronger compared to Dirac SS,[31] the coherence time in the $Bi_2Se_3$ bulk intermediate state becomes long enough, thus significantly diminishing the probability of two-photon excitation. This behavior seems to have the same origin as the Franck–Condon blockade in a suspended nanostructure.[27] For the same reason, the excitation of carriers from the valence band (VB) to the CB is expected to occur within one-photon pumping, since an off-resonant vertical transition is involved in the presence of strong electron-phonon coupling in the $Bi_2Se_3$ bulk.

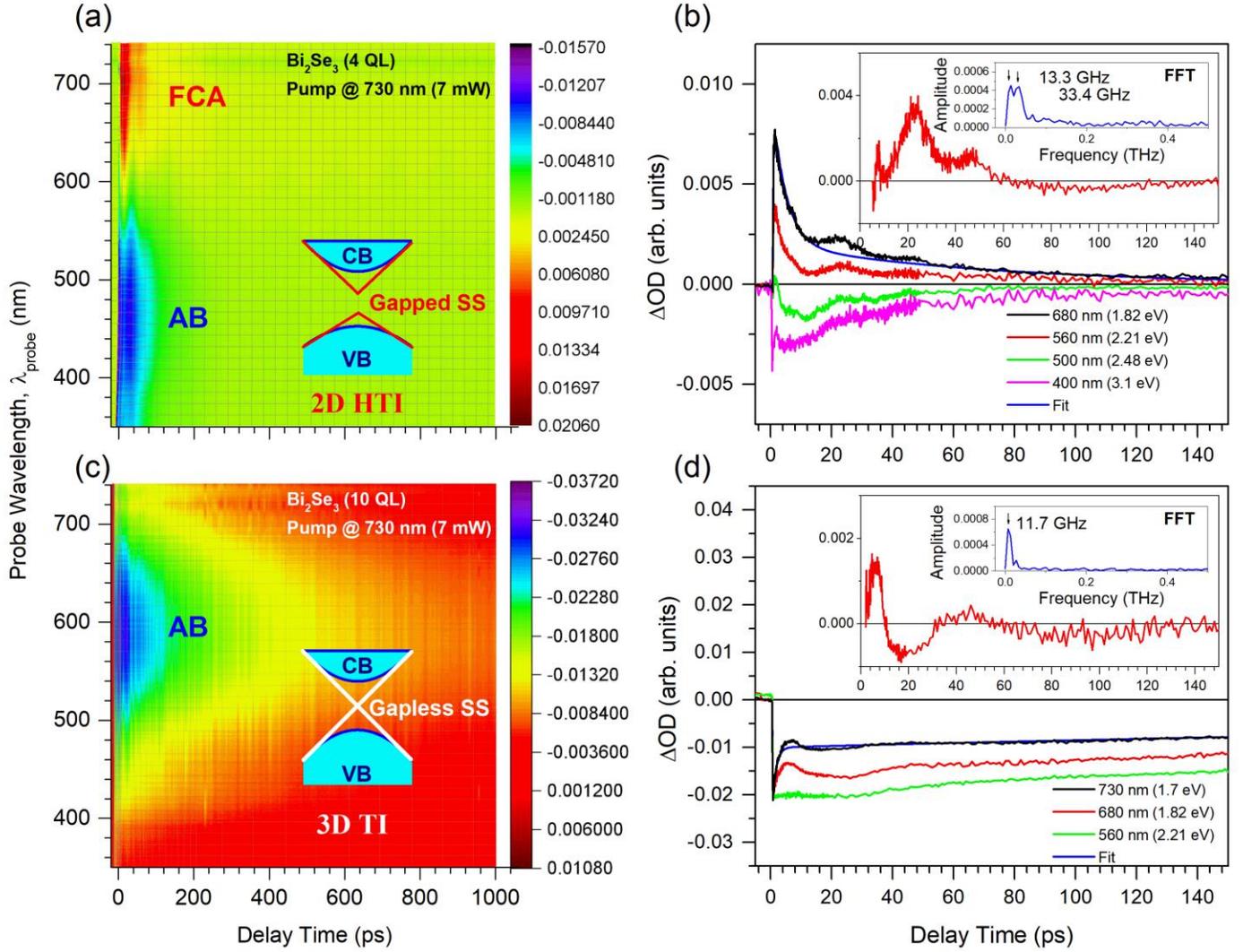

**Fig. 2.** (a) and (c) Representative pseudo-color TA spectra plots of ~4 QL and ~10 QL thick Bi$_2$Se$_3$ films, respectively, which were measured with the ~730 nm pumping (~1.7 eV photon energy) of ~7.0 mW power. The color bar is shown to the right. The spectral ranges associated with absorption bleaching (AB) and free carrier absorption (FCA) are indicated. The schematic representation of two topological phases, corresponding to the 2D HTI (gapped SS) and the 3D TI (gapless SS) are shown. (b) and (d) The corresponding pump–probe traces probed at the specific wavelengths, as indicated by the corresponding colors. The blue curve shows the best fit to the data. Insets represent the extracted oscillatory part and the corresponding fast Fourier transformation (FFT) with the center frequency, as indicated.

The TA spectra plot of the 4 QL thick film that is usually referred to as the 2D HTI phase (gapped SS)[4,5] demonstrates the positive and negative contributions with a short decay time within ~150 ps [Fig. 2(a)]. In contrast, the TA spectra plot of the 10 QL thick film that is known as 3D TI phase (gapless SS) is characterized exclusively by a negative contribution with a much longer decay time exceeding ~1.0 ns. As we mentioned in the Introduction, there have been numerous studies on ultrafast carrier dynamics in $Bi_2Se_3$ films, in which the most common pump-probe reflection configurations employing the identical wavelength pumping and probing beams of a Ti:Sapphire laser were applied.[13-18] The resulting pump-probe responses were always negative, indicating that their origin is being associated with absorption bleaching (AB) (Pauli blocking). In general, the AB response magnitude in semiconductors is supposed to be weakly dependent on the photoexcited carrier density, scaling as $n^{1/6}$.[25] The negative contribution of the TA spectra plot of the 3D TI phase [Fig. 2(c)] and the corresponding negative pump-probe traces measured at different photon energies [Fig. 2(d)] demonstrate hence a typical behavior of the AB response. Moreover, if the transiently excited electron population is probed at ~1.7 eV photon energy, one can recognize a typical two-stage decay. Additionally, there is a one-cycle oscillatory behavior, the estimated frequency of which (11.7 GHz) [Fig. 2(d), Inset] is well consistent with the acoustic phonon dynamics previously reported elsewhere.[13,18] The initial fast decay stage within a few picoseconds is usually associated with the cooling of hot electrons through the Fröhlich interaction mechanism.[6,7,13-18] This decay is known to shorten with decreasing film thickness, approaching typical values of noble metals and thus indicating a dominant effect of Dirac SS on the carrier relaxation dynamics in ultrathin $Bi_2Se_3$ films.[16] A much longer decay time characterizing the slow decay stage is also film thickness dependent and results from the interplay of the two relaxation channels. The first of them deals with a metastable population of the upper Dirac cone of SS2 by relaxing electrons and their radiative/non-radiative recombination with holes residing in the lower Dirac cone of SS1 (~5.0 – 150 ps) (Fig. 1).[17] The second relaxation channel involves further LO-phonon-assisted relaxation via the Dirac point of SS2 and a metastable population of the CB edge (≤3.5 ps), continuously feeding the upper Dirac cone of SS1 (Fig. 1). Finally, radiative/non-radiative recombination of electrons and holes residing in the upper and lower Dirac cones of SS1, respectively, governs the long-time carrier relaxation within a few nanoseconds.[13-18] As the film thickness decreases, the first relaxation channel begins to dominate over the second channel, thus shortening the overall decay time of the slow decay stage.[17] It is noteworthy that the final slow decay stage for both relaxation channels involves Dirac SS, where carriers localize prior to recombination, as it is required for energy and momentum conservation.[32]

We also note that the two-stage decay dynamics has been observed in numerous experiments performed in the visible and mid-IR/THz regions,[13-19,22] although the visible range probing monitors mainly the higher energy bulk states and SS2, whereas the IR/THz probing deals with the lower energy bulk states and SS1. However, as we report here, this two-stage decay and the one-cycle oscillatory behavior are not necessarily unique, since both disappear completely upon increasing the probing photon energy, thus leading to the single stage slow (almost stable within first 20 ps) and smooth relaxation trend [Fig. 2(d)]. As we show farther below, this behavior is caused by a significant weakening of electron-phonon interaction when switching from the bulk state relaxation dynamics toward the SS relaxation dynamics.

The relaxation of two-photon-excited carriers in the 2D HTI phase seems to be more complicated [Fig. 2(b)]. Specifically, the initial fast decay stage of the AB response becomes extremely short (~0.2 ps), being hence comparable to the AB response rise time of ~0.15 ps. This behavior is believed to originate from the overlap of the negative and positive contributions. Correspondingly, because the positive response rises slightly slower than the AB response, their overlap forms a narrow negative feature, the intensity of which gradually decreases with decreasing probing photon energy [Fig. 2(b)]. Similarly, the one-cycle oscillatory behavior of the AB response overlaps with the higher frequency oscillations of the positive response (33.4 GHz), which can be associated with the coherent Dirac plasmon.[8] These two oscillatory behaviors can be recognized exclusively using the fast Fourier transformation (FFT) procedure [Fig. 2(b), Inset].

It is worth noting that the positive contribution to the transient reflectivity traces has never been reported for $Bi_2Se_3$ when the visible range pump-probe configurations were applied. Nevertheless, the positive amplitude pump-probe response from semiconductors, which generally characterizes the photoinduced absorption regime, is known to result from the many-body effects such as bandgap renormalization and FCA.[25] Because the plasma frequency for carrier densities usually photoexcited in $Bi_2Se_3$ typically has values in the IR/mid-IR region, to apply the FCA concept to the visible region, collisions between free carriers and the lattice ions should be considered. This kind of FCA is also known as the inverse bremsstrahlung absorption, the process that allows energy and momentum to be conserved simultaneously when an electron absorbs a light quantum.[25] On the contrary, FCA in the mid-IR/THz range is due to more common Drude type absorption, nevertheless, also providing a positive contribution to the transient reflectivity and transient conductivity traces.[20,22]

In high-quality molecular beam epitaxy-grown films of $Bi_2Se_3$, bandgap renormalization is expected to be mainly governed by interaction with LO-phonons (polaronic bandgap renormalization),[25,33] whereas FCA should occur when the local density of carriers becomes high enough owing to their accumulation in some states. Using the pumping power dependences of the pump-probe trace peak intensity, one can set the bandgap renormalization and FCA processes apart, since they are differently dependent on the photoexcited carrier density ($n^{1/3}$ versus $n$, respectively).[25] As discussed further below, because of the linear pumping power dependence of the positive response peak intensity, the positive contribution can be attributed to the inverse bremsstrahlung type FCA. Specifically, this kind of FCA is expected to appear in TA spectra with some delay after excitation, since an accumulation of free carriers is required. Alternatively, the polaronic bandgap renormalization, if any, seems to be more efficient in the $Bi_2Se_3$ bulk structure due to much stronger electron-phonon coupling compared to that in Dirac SS.[30,31]

To summarize, starting immediately after excitation, the AB response images the transiently excited carrier population in the occupied Dirac SS, as well as in the occupied bulk states of the CB and VB (CB-AB and VB-AB, respectively).[25] In contrast, the FCA response monitors the same transiently excited carrier population in occupied states, however, when the local carrier density becomes high enough to allow carriers to collide with the lattice ions and hence to absorb probing light through the inverse bremsstrahlung type FCA mechanism. This electron density effect is expected to be caused by collisions between free electrons residing in Dirac SS and the lattice ions of the $Bi_2Se_3$ bulk. It is important to note that both the AB and FCA effects act simultaneously and their overlap is appeared as a redistribution between the corresponding transient response amplitudes, depending on the local carrier density. Because there is no way to reach the same local carrier density in the $Bi_2Se_3$ bulk, the inverse bremsstrahlung type FCA mechanism becomes Dirac SS sensitive. Consequently, the positive contribution of TA spectra observed in the 2D HTI phase (4 QL) suggests that in this sample the density of relaxing electrons upon their accumulation in SS2 with energies around ~1.8 eV becomes much higher compared to that in SS2 of the 3D TI phase (10 QL) [Fig. 2(a) and (c)]. Such a crucial change in the ultrafast carrier relaxation dynamics shown in Fig. 2 for ultrathin $Bi_2Se_3$ films can be associated with the topological phase transition. The schematic representation of the samples of various topological phases used in this study and the corresponding predicted shapes of TA spectra are shown in Fig. 3.

Because the amplitude of both the negative and positive transient responses are controlled by the local density of relaxing electrons, the specific shape of TA spectra can be predicted using a traditional presentation of the electron density as a product of the density of states (DOS) [$g(E)$] and the Fermi-Dirac function [$f(E)$], $n = \int g(E)f(E)dE$. The 3D TI phase is characterized by a combination of the bulk states and gapless SS with the square root type and linear type $g(E)$, respectively [Fig. 3(a) and (b)].[32,34,35] Alternatively, the 2D HTI phase deals with a combination of the bulk states and gapped SS with the square root type and constant type $g(E)$, respectively,[31] because there are no gapless SS with topological protection and gapped SS can be considered as the actual quantum well discrete states. Similarly, $g(E)$ is the constant type for the 2D TTI phase because quantum confinement mainly governs the properties of gapped SS [Fig. 3(a) and (b)].[5,36] The resulting densities of electrons versus their energies in occupied states (nearly versus the probe photon energy in our experiments) are shown in

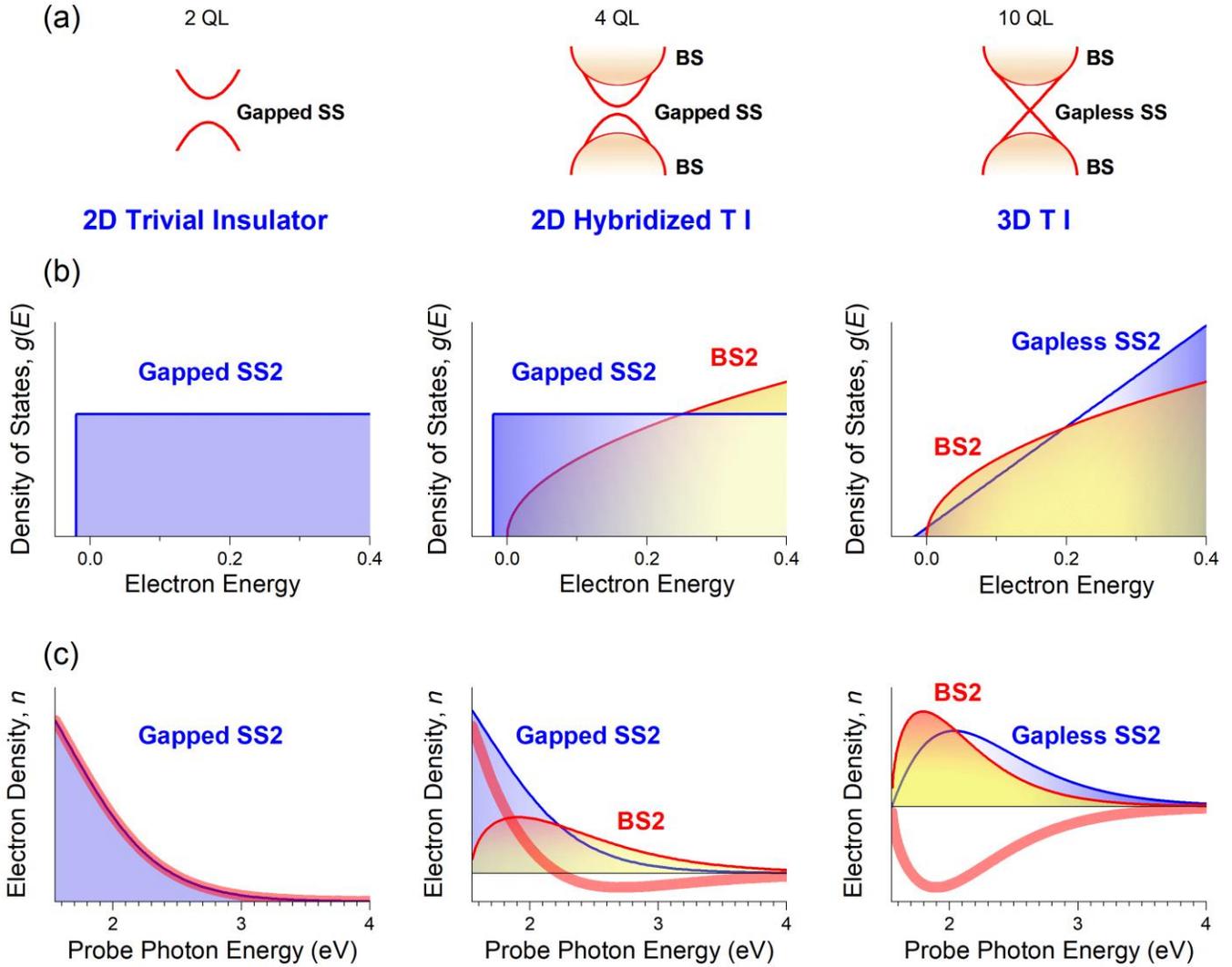

**Fig. 3.** (a) Band diagrams for the various topological phases of $Bi_2Se_3$ ultrathin films of different thicknesses, as indicated. (b) The corresponding density of states (DOS) for the bulk states (BS2) and SS2. (c) The resulting electron density distribution for BS2 and SS2 as a function of electron energy (nearly equal to the probe photon energy). The wide light-red curves represent the predicted spectral shapes of the corresponding TA spectra.

Fig. 3(c) together with the predicted spectral shapes of TA spectra. Specifically, in the 2D TTI phase, the TA spectrum is expected to be positive when completely ignoring the bulk states effect and if the density of relaxing electrons in the gapped SS becomes high enough. The corresponding spectral shape is hence being determined by the high energy tail of the Fermi-Dirac distribution. In the 2D HTI phase, in addition to the previous case, a part of photoexcited electrons may occupy the bulk states, giving rise hence to the negative contribution associated with absorption bleaching. The resulting spectral shape is determined by a sum of positive and negative contributions and therefore the TA spectrum changes its sign at some characteristic crossing point. Finally, in the 3D TI phase, the density of relaxing electrons in the occupied bulk states and gapless SS is not high enough to appear through the inverse bremsstrahlung type FCA and therefore only the negative contributions associated with absorption bleaching can be observed. However, as the relative electron density in the bulk states and SS redistributes within the carrier relaxation process, the spectral shape of the TA spectrum is expected to vary since different types of DOS characterize the bulk states and Dirac SS [Fig. 3(b)].

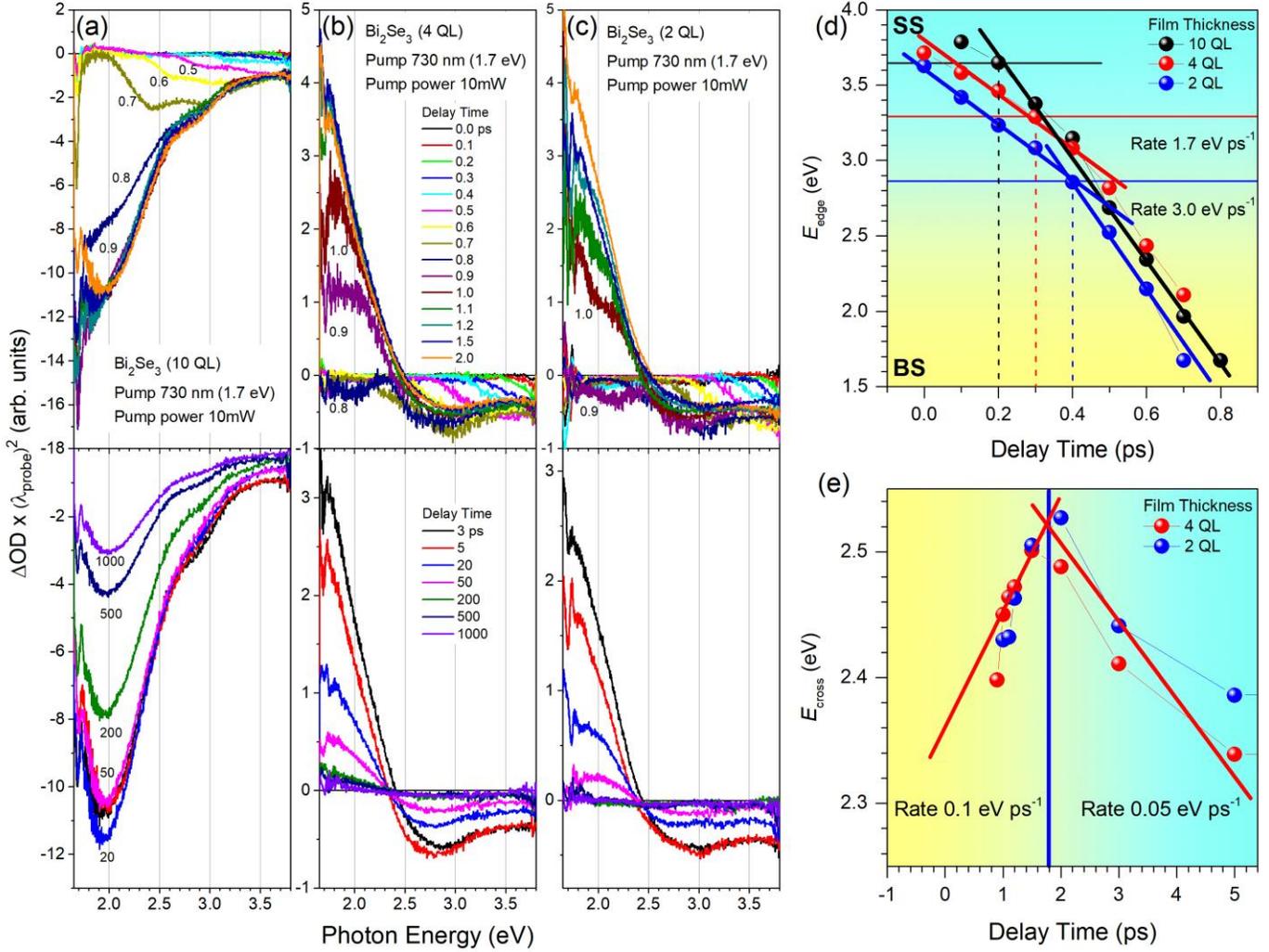

**Fig. 4.** (a), (b), and (c) A set of TA spectra of $Bi_2Se_3$ films with various thicknesses of 10 QL, 4 QL, and 2 QL, respectively. The spectra were measured using the ~730 nm pumping (1.7 eV photon energy) of ~10 mW power at delay times, as indicated by the corresponding colors. The numbers at some of the spectra duplicate the actual delay times in picoseconds for clarity. (d) and (e) The temporal evolution of the low-energy edge of transiently excited electron population ($E_{edge}$) and the crossing point energy ($E_{cross}$), respectively [both being explicitly defined in Fig. 5(b)], for $Bi_2Se_3$ films of different thicknesses, as indicated by the corresponding colors. The linear fits (the corresponding color straight lines) and the corresponding electron energy loss rates in eV $ps^{-1}$ units are shown for the higher energy Dirac SS and the bulk states (BS), as indicated. The corresponding color horizontal solid lines and the vertical dashed lines in (d) indicate the energy and time ranges, respectively, where the electron energy loss rate changes.

Figure 4(a), (b), and (c) and Figures 5(a) and 6(a) show a snapshot TA spectral imaging measured for the three samples corresponding to the discussed three topological phases. The TA spectra mainly follow all tendencies presented in Fig. 3(c). First, one should distinguish between two relaxation dynamics associated with VB-AB (a

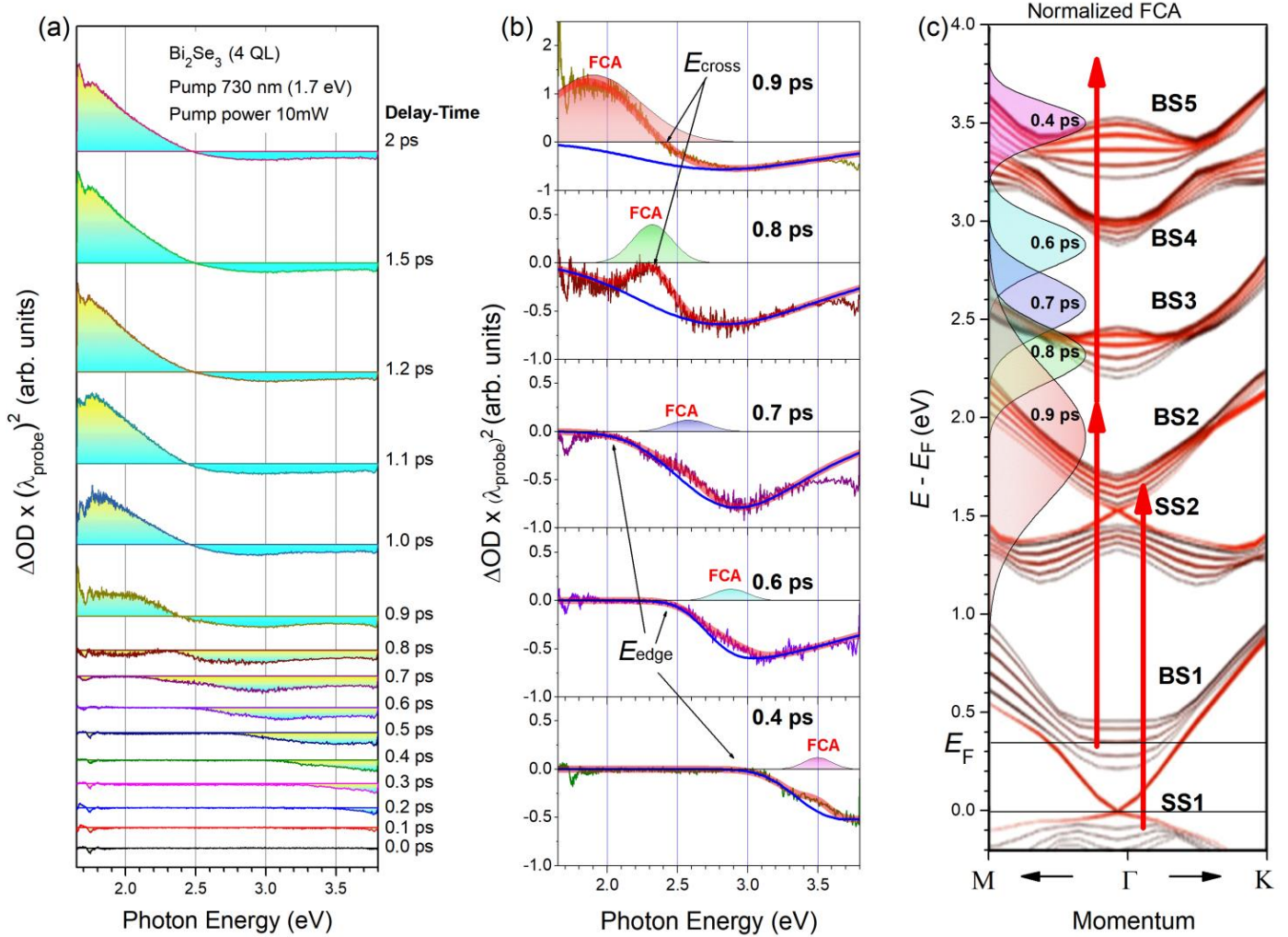

**Fig. 5.** (a) A set of TA spectra of the 4 QL thick $Bi_2Se_3$ film measured within the 0.0 – 2.0 ps timescale using the ~730 nm pumping (~1.7 eV photon energy) of ~10 mW power [the same as shown in Fig. 4(b)]. The zero-intensity lines of TA spectra were shifted along the ΔOD axis for better observation. (b) Some TA spectra shown in (a) and the best fit to the data (light-red broad curves). The absorption bleaching (AB) and free carrier absorption (FCA) contributions (the blue color curves and filled curves, respectively) were separated, as illustrated. The low-energy edge of the transiently excited electron population ($E_{edge}$) and the crossing point energy ($E_{cross}$) are indicated. (c) The band structure of the 6 QL thick $Bi_2Se_3$ film calculated in Ref. 3 and the one-photon and two-photon pumping transitions (red vertical up arrows) originating from the VB states and from the upper Dirac cone of SS1 below the Fermi energy ($E_F$), respectively. The bulk states and the Dirac surface states are marked as BS and SS, respectively. The normalized FCA contributions extracted in (b) are shown for specific delay times, as indicated.

narrow negative feature exactly matching with the pumping photon energy) and CB-AB (the broad negative feature covering the entire visible region for the 3D TI phase and a part of the visible region for the 2D HTI and 2D TTI phases). The negative VB-AB feature peaked at ~1.7 eV is constantly presented in TA spectra of all the topological phases irrespective of delay time and therefore it can be associated with the one-photon pumping of valence electrons to the CB [Fig. 5(c) and Fig. 6(c)]. The photoexcited holes in the VB rapidly relax down within the ~0.1 ps timescale,[16] occupying the VB edge states and the lower Dirac cone of SS1 for a long time comparable to the inverse repetition rate of the laser used for measurements (~1.0 ms). Consequently, the accumulated holes block out the probing optical transitions originating from the mentioned states, as long as the sample is being optically pumped, the process that appears as a VB-AB response.[8,25] Because this feature completely disappears when the probing beam is blocked, it has nothing to do with the scattered pumping light acting only within the truly short timescale of ~0.1 ps.[25]

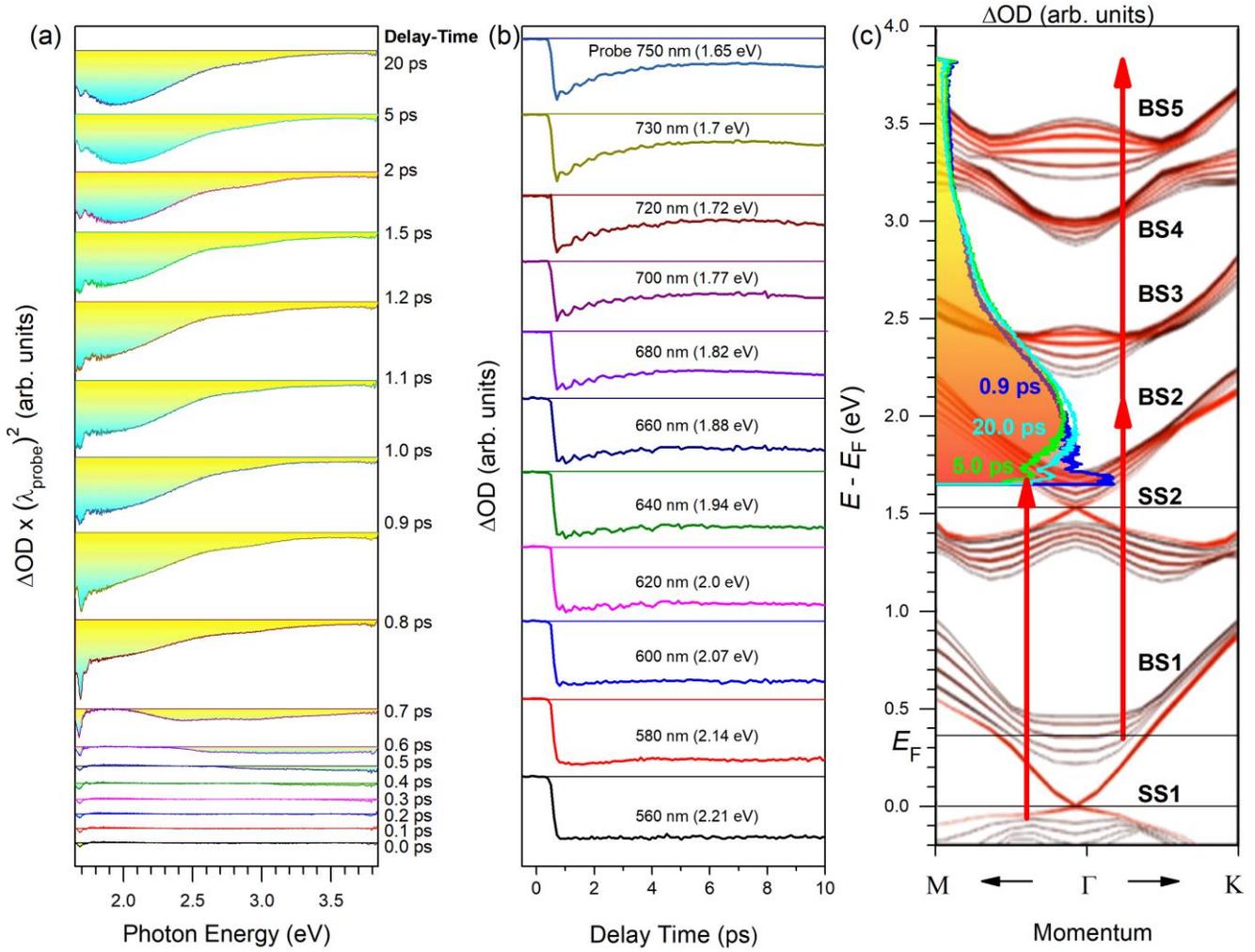

**Fig. 6.** (a) A set of TA spectra of the 10 QL thick $Bi_2Se_3$ film measured within the 20 ps timescale using the ~730 nm pumping (~1.7 eV photon energy) of ~10 mW power [the same as shown in Fig. 4(a)]. The zero-intensity lines of TA spectra were shifted along the ΔOD axis for better observation. (b) The corresponding pump–probe traces measured at the specific probing wavelengths, as indicated for each of the traces. (c) The band structure of the 6 QL thick $Bi_2Se_3$ film calculated in Ref. 3 and the one-photon and two-photon pumping transitions (red vertical up arrows) originating from the VB states and from the upper Dirac cone of SS1 below the Fermi energy ($E_F$), respectively. The bulk states and the Dirac surface states are marked as BS and SS, respectively. The TA spectra from (a) measured at 0.9 ps, 5.0 ps, and 20 ps delay times are shown.

On the contrary, as we discussed above, the CB-AB feature is associated with two-photon pumping [Fig. 5(c) and Fig. 6(c)]. For the 3D TI phase, the corresponding AB response develops within the ~0.9 ps timescale, gradually occupying the entire visible region by extending from the initial position at ~3.8 eV toward ~1.6 eV and finally being peaked at ~2.0 eV [Figs. 4(a) and 6(a)]. This relaxation dynamics demonstrates a spectacular cooling of two-photon-excited electrons toward the Dirac point of SS2 [Fig. 6(a) and (c)], followed by stabilization within first ~20 ps after excitation (this stabilization corresponds to what is showing in Fig. 2(d) for the corresponding pump-probe traces) and final decay within a few nanoseconds [Fig. 4(a)]. The stabilization of the AB response amplitude at ~2.0 eV for a long time confirms the dynamical accumulation of electrons in the upper Dirac cone of SS2. However, as we mentioned above, the electron density remains to be not high enough to appear through the inverse bremsstrahlung type FCA. The dynamical accumulation of electrons in SS2 observed for the 3D TI phase is also well consistent with a higher nonequilibrium mobility found in SS2 compared to equilibrium conductivity in SS1.[23] We also note that the discussed electron cooling dynamics unambiguously states that the rise time of the pump-probe traces caused by absorption bleaching, which is usually being associated with electron-electron thermalization (~0.2 - 0.3 ps) when the identical wavelength pumping and probing beams were applied,[8,13-18] rather characterizes the rate of electron energy relaxation through the LO-phonon cascade emission than the electron-electron thermalization process itself.

For the 2D HTI and 2D TTI phases, the electron relaxation dynamics appearing through TA spectra begins initially with the development of the AB contribution as well. However, the positive contribution is emerged at ~0.9 ps after excitation, gradually dominating the TA spectrum and being peaked in the energy range slightly above the Dirac point energy of SS2 (~1.5 eV) within ~2.0 ps, followed by the corresponding decay within ~200 ps [Fig. 4(b) and (c) and Fig. 5(a)]. The time at which the positive contribution of TA spectra is maximized in the 2D HTI and 2D TTI phases (~2.0 ps) is longer than that required for the negative AB contribution to be developed in the 3D TI phase (~0.9 ps), despite nearly identical peak position of the opposite sign contributions. This behavior clearly demonstrates the accumulation of electrons in the upper Dirac cone of SS2 for the 2D HTI and 2D TTI phases, the process that appears through the inverse bremsstrahlung type FCA. We associate this behavior with the progressive emergence of strong coupling between SS2 on the opposite surfaces of the film with decreasing its thickness,[4] which significantly redistributes relaxing electrons toward SS2 and hence increases their density. It is worth noting that the positive FCA response at shorter than ~0.9 ps delay times can also be extracted from the negative AB response if an accurate enough fitting procedure is applied [Fig. 5(b)]. The resulting dynamics of the FCA response [Fig. 5(c)] monitors hence directly the relaxation of the higher density part of the transiently excited electron population, which we associate with that dealing with the cooling of electrons in the higher energy SS. This finding suggests that both the AB and FCA responses enable monitoring the ultrafast cooling of photoexcited electron-hole plasma in ultrathin $Bi_2Se_3$ films.

By comparing to the 3D TI phase, we conclude that the carrier relaxation dynamics is exclusively influenced by whether the Dirac point is presented between the upper and lower Dirac cones of SS2. Consequently, the accumulation of carriers in SS2 and the corresponding FCA response magnitude should be considered in terms of carrier relaxation through the gapless SS2 or through the gapped SS2. Because the upper and lower Dirac cones of SS2 are initially unoccupied, the highly energetic electrons relaxing through the LO-phonon cascade emission and gradually occupying the upper Dirac cone can scatter into the lower Dirac cone through the LO-phonon-assisted or disorder-assisted mechanisms,[37] despite zero density of states in the Dirac point. On the contrary, this kind of scattering is substantially slowed down when the gap between the Dirac cones exceeds the LO-phonon and resonant defects energies. This conclusion is well consistent with the maximal bulk polar phonon energy in $Bi_2Se_3$ of 21.4 meV[28] and the SS1 gap opening that was measured using ARPES and ranges 41 meV, 70 meV, 138 meV, and 252 meV for the 5 QL, 4 QL, 3 QL, and 2 QL thick films, respectively.[4] Here we reasonably assumed that the gap opening range in SS2 is comparable to that in SS1. Consequently, once electrons in the 2D HTI and 2D TTI phases progressively accumulate in the upper Dirac cone of the gapped SS2, their density significantly increases, giving rise to the positive contribution in TA spectra associated with the inverse-bremsstrahlung type FCA. The SS2 act hence as a valve substantially controlling the rate of electron relaxation in ultrathin $Bi_2Se_3$ films.

Although the carrier relaxation dynamics in both the 2D HTI and 2D TTI phases seems to be quite similar, one can recognize that the positive contribution in the 2D TTI phase is more extended toward the higher energy range compared to that in the 2D HTI phase [Fig. 4(b) and (c)]. This behavior of the 2D TTI phase points hence to the larger population of relaxing electrons occupying more energetic SS2. Another important observation is that the initial relaxation of two-photon excited electrons from the higher energy SS of both the 2D HTI and 2D TTI phases involves the bulk states. However, the bulk states effect in the 2D TTI phase is much weaker, nevertheless there still exists. This behavior suggests that the bulk states still noticeably affect the initial relaxation dynamics of two-photon-excited electrons even in the 2 QL thick $Bi_2Se_3$ film, thus improving the ideal model of 2D TTI phase presented in Fig. 3. Furthermore, we note that the AB and FCA responses in the 2D HTI and 2D TTI phases reveal the much shorter decay times compared to that observed for the negative response in the 3D TI phase (Fig. 2). This behavior suggests that the electron relaxation channel associated with a metastable population of the CB edge and the upper Dirac cone of SS1 (the second relaxation channel in the 3D TI phase) is completely suppressed. Consequently, another relaxation channel dealing with radiative/non-radiative recombination of electrons residing in the upper Dirac cone of SS2 and holes residing in the lower Dirac cone of SS1 (the first relaxation channel in the 3D TI phase) dominates the relaxation dynamics.[17] We also note that the suppression of the second relaxation channel in the 2D HTI and 2D TTI phases creates great conditions required to achieve population inversion and recombination lasing.

**3.2 The electron relaxation and accumulation rates.**

The temporal evolution of TA spectra shown in Fig. 4(a) (b), and (c) allowed us to estimate the electron relaxation rates in all the topological phases. Specifically, we used the shift of the low-energy edge of the transiently excited electron population ($E_{\text{edge}}$) with delay time within ~1 ps [Fig. 5(b)]. The resulting temporal dynamics is quite similar for all the topological phases and clearly demonstrates the existence of the two-stage electron cooling process [Fig. 4(d)]. We associate this two-stage behavior with the relaxation of two-photon-excited electrons through the LO-phonon cascade emission occurring initially in the higher energy SS and switching subsequently to the higher energy bulk states [Fig. 5(c) and Fig. 6(c)]. The switching between two dynamics occurs at ~0.2 ps, ~0.3 ps, and ~0.4 ps for the 10 QL, 4 QL, and 2 QL thick films, respectively. The longer time of the initial cooling dynamics in the higher energy SS for the thinner films is governed by stronger coupling between the opposite surface Dirac SS2. The corresponding rates are ~1.7 eV ps$^{-1}$ for the Dirac SS and ~3.0 eV ps$^{-1}$ for the bulk states [Fig. 4(d)]. The about two times slower cooling rate of electrons in SS compared to the bulk states can be attributed to much weaker electron-phonon coupling in Dirac SS. In general, the relaxation of photoexcited electrons is associated with Fröhlich interaction and can be treated using the rate of emission of phonons by hot electrons,[16]

$$\frac{1}{\tau_{e-ph}} = \frac{e^2}{4\pi\varepsilon_0 \hbar}\left(\frac{2m_e^*\hbar\omega_{ph}}{\hbar^2}\right)^{1/2}\left(\frac{1}{\varepsilon_\infty} - \frac{1}{\varepsilon_s}\right), \tag{1}$$

where $\tau_{e-ph}$ is the electron-phonon inelastic scattering time, $e$ is the electron charge, $\varepsilon_0$ is the permittivity of free space, $\hbar\omega_{ph}$ is the phonon energy, $m_e^* = 0.14\, m_0$ is the electron effective mass in the $\Gamma$ valley ($m_0$ is the free-electron mass), and $\varepsilon_\infty = 9$ and $\varepsilon_s = 100$ are the high-frequency and static dielectric constants, respectively. The resulting scattering times ($\tau_{e-ph}$) and the corresponding electron energy relaxation rates in the Bi$_2$Se$_3$ bulk for all polar optical modes[28] are listed in the Table 1.

| Polar optical mode | Raman frequency (cm$^{-1}$) | $\hbar\omega_{ph}$ (meV) | $\tau_{e-ph}$ (fs) | Electron relaxation rate (eV ps$^{-1}$) |
|---|---|---|---|---|
| $A_{1g}^1$ | 73 | 9.0 | 30.7 | 0.29 |
| $E_g^2$ | 130 | 16.1 | 22.9 | 0.7 |
| $A_{1g}^2$ | 173 | 21.4 | 19.9 | 1.08 |

Table 1. Electron relaxation rates due to Fröhlich interaction.

One can see that the electron relaxation rate varies significantly for different bulk polar phonon modes, however, it always remains much smaller than those experimentally observed. Because the phonon softening effect occurring in SS1 is known to be less than 10%,[6] it cannot be the reason for such a big difference. Consequently, the multiphonon relaxation mechanism should be taken into consideration to meet the measured electron energy relaxation rates. We note that the coupling between phonon modes in ultrathin Bi$_2$Se$_3$ films seems to be natural and the combined $E_g^2 + A_{1g}^2$ phonon mode has been detected by Raman spectroscopy.[28]

The accumulation of two-photon-excited electrons in the upper Dirac cone of SS2 for the 2D HTI and 2D TTI phases can also be characterized by the corresponding rate. Specifically, as a measure of electron population accumulated in the upper Dirac cone of SS2, we consider the characteristic energy of the crossing point in TA spectra where the negative (AB) trend switches to the positive (FCA) trend, $E_{\text{cross}}$ [Fig. 5(b)]. Consequently, the position of this crossing point is determined by the number of electrons accumulated in SS2, whereas the shift of this point with delay time would allow the accumulation rate to be estimated. Figure 4(e) demonstrates that the number of accumulated electrons rises sharply within the ~0.9 – 1.9 ps timescale, followed by a fast decrease within ~5 ps and a much slower decrease accompanying the overall decay of the TA response. The estimated rate of the electron accumulation in the upper cone of SS2 is ~0.1 eV ps$^{-1}$, thus being much lower than the rate of the electron energy relaxation in the higher energy SS and the bulk states. This behavior unambiguously proves that the electron accumulation process is accompanied by recombination of electrons accumulated in the upper Dirac cone of SS2 and holes residing in the lower Dirac cone of SS1. As we mentioned above, this recombination process in the 2D HTI and 2D TTI phases becomes dominant due to the suppression of the second relaxation channel in SS2 when the gap between the Dirac cones is formed. The slowdown in electron accumulation dynamics is hence controlled by the electron-hole recombination rate.

**3.3 The LO-phonon-assisted vertical transport and electron-phonon coupling in Dirac SS2.**

One of the most important findings of this study is related to the electron-phonon coupling in the bulk states and Dirac SS. Figure 6(b) shows the pump-probe traces measured for the 3D TI phase as a function of probing wavelength. One can see that once the probing wavelength decreases (the probing photon energy increases), in addition to the loss of the fast decay stage discussed above [Fig. 2(d)], the oscillatory behavior associated with the LO-phonon mode also disappears when the probing photon energy approaches ~2.0 eV. The frequency of coherent LO-phonon oscillations (~2.2 THz) closely matches with those previously reported for the $Bi_2Se_3$ bulk structures [Fig. 7(a)].[6,13,15,16] To explain this remarkable finding associated with the oscillatory behavior loss in the pump-probe traces, TA spectra measured at ~0.9 ps, ~5.0 ps, and ~20.0 ps delay times can be used [Fig. 6(a) and (c)]. Specifically, the intensity of the TA spectrum ranging from ~1.65 eV to ~2.0 eV drops down significantly within ~5.0 ps, whereas it remains almost unchanged when electrons occupy states with energies higher than ~2.1 eV. The discussed temporal dynamics observed in TA spectra at different delay times corresponds to the loss of the fast decay stage and coherent LO-phonon oscillations observed in the pump-probe traces [Fig. 2(d), Fig. 6(b), and Fig. 7(a)]. Furthermore, this observation suggests the existence of two transiently excited populations of electrons occupying different spatially separated states in the 3D TI phase. The first population appears to peak near ~2.0 eV and dominates at longer delay times, whereas the second one peaked near ~1.7 eV is maximized within the sub-picosecond timescale (~0.9 ps), subsequently decaying within ~5.0 ps, slightly recovering within ~20 ps, and finally decaying within a longer time exceeding ~1.0 ns [Fig. 4(a)]. This behavior of the TA spectrum is well consistent with the one-cycle oscillatory trend observed in the corresponding pump-probe traces shown in Fig. 2(d) and Fig. 7(a). As we mentioned above, the one-cycle oscillatory trend in the 3D TI phase has been observed in numerous ultrafast experiments employing the most common pump-probe configuration using the identical wavelength pumping and probing beams of a commercial ultrafast Ti:Sapphire laser.[13-18] Although this one-cycle oscillatory behavior, in general, has been attributed to acoustic phonons, the application of the surface sensitive technique (pump-probe SHG) allowed to recognize the SS-bulk-SS vertical electron transport that provides similar one-cycle oscillatory trend and that has been found to be responsible for the excitation of coherent acoustic Dirac plasmons.[8] Consequently, the transiently excited electron population peaked in TA spectra of the 3D TI phase at ~1.7 eV can be attributed to the bulk states (BS2) [Fig. 6(c)]. A decrease in the amplitude of this electron population through the LO-phonon cascade emission within ~5.0 ps indicates hence the LO-phonon-assisted vertical transport of electrons from the bulk states (BS2) toward SS2. Such a redistribution of electrons between BS2 and SS2 allows the transiently excited electron population peaked in TA spectra near ~2.0 eV and associated with SS2 to be observed explicitly [Fig. 6(c)]. It is clear now that the corresponding loss of coherent LO-phonon oscillations in pump-probe traces with increasing the probing photon energy toward ~2.0 eV results from the coherence length shortening when switching from the bulk states relaxation dynamics toward the SS relaxation dynamics. We associate this switching with different types of electron-phonon coupling in the bulk states and Dirac SS (inelastic and quasi-elastic, respectively).[38]

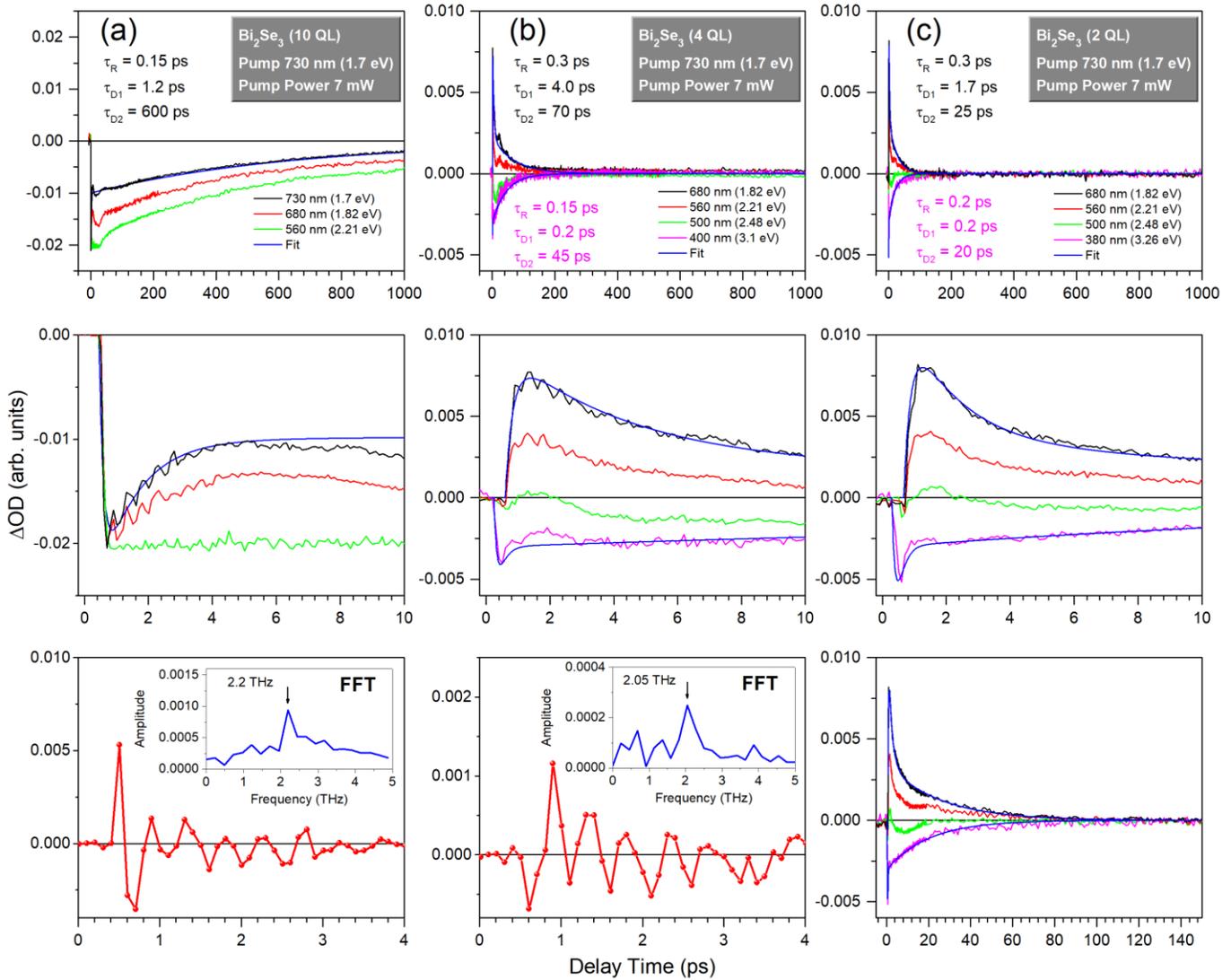

**Fig. 7.** (a), (b), and (c) A set of pump–probe traces for Bi$_2$Se$_3$ films with various thicknesses of 10 QL, 4 QL, and 2 QL, respectively. The traces were measured using the ~730 nm pumping (~1.7 eV photon energy) of ~7.0 mW power and at different probing wavelengths, as indicated by the corresponding colors. Each of columns presents the same traces, but of different delay time scales. The traces in (a) and (b) are the same as those shown in Fig. 2(d) and (b), respectively. The blue curve shows the best fit to the data. The results of the fit are listed in the upper panels. The bottom panels in (a) and (b) represent the extracted oscillatory part and the corresponding fast Fourier transformation (FFT) with the center frequency, as indicated.

The electron redistribution between the bulk states (BS2) and SS2 in the 3D TI phase is expected to occur due to different types of DOS, as shown in Fig. 3(b) and (c). Specifically, the electron population is two-photon-excited initially in the higher energy SS. Further relaxation of electrons through the LO-phonon cascade emission first occurs via the higher energy SS, switching afterward to the higher energy bulk states at ~0.2 - ~0.4 ps (as discussed in the preceding section), and finally populating the bulk states (BS2) within ~0.9 ps. The subsequent LO-phonon-assisted vertical transport redistributes electrons back toward SS2 within ~5.0 ps, because of the higher DOS in SS2 for more energetic electrons [Fig. 3(b)]. The subsequent recoil effect, partially moving electrons back to the bulk states within the ~20 ps timescale and completing the one-cycle oscillatory dynamics, occurs because the rate of LO-phonon-assisted relaxation via the Dirac point of SS2 is not as high as that of electron relaxation in the higher energy bulk states and Dirac SS.

The LO-phonon-assisted relaxation dynamics in the 2D HTI phase differs significantly [Fig. 7(b)]. The remaining narrow feature, which is associated with the fast decay stage in the AB response of the 3D TI phase, in the 2D HTI phase tends to disappear with decreasing probing wavelength, thus indicating that it is due to the overlap of the negative and positive contributions. Additionally, the coherent LO-phonon oscillatory behavior in the AB response of the 2D HTI phase disappears completely, thus suggesting that the LO-phonon-assisted vertical electron transport becomes negligible due to strong coupling between SS2 on the opposite surfaces of the ultrathin $Bi_2Se_3$ film. In contrast, the coherent LO-phonon oscillatory behavior begins appearing in the FCA response, but with lower frequency of ~2.05 THz. The latter frequency exactly matches with that observed for SS1 using the TrARPES technique and attributed to the softening of the bulk LO-phonon mode near the crystal surface.[6] Consequently, the appearance of coherent LO-phonon oscillations in the FCA response seems to be related to the LO-phonon mode softening, which elongate the coherence length of electron oscillations involved into quasi-elastic electron-phonon interaction in SS2. The latter behavior is known to appear in the image-potential states at the metal surfaces[38] and is well consistent with the collisional nature of the inverse bremsstrahlung type FCA in SS2. This observation also suggests that the main electron relaxation mechanism in Dirac SS of the 3D TI phase is governed by the interaction of massless fermions with polar phonons of the $Bi_2Se_3$ bulk, despite the quasi-elastic character of electron-phonon interaction.

Another remarkable finding is that the coherent LO-phonon oscillations disappear completely in the 2D TTI phase for both the negative and positive contributions [Fig. 7(c)]. This behavior demonstrates a result of shortening coherence length due to the preferably collisional (elastic) interaction of electrons accumulated in the upper Dirac cone right above the SS2 gap with the $Bi_2Se_3$ bulk. This kind of purely elastic electron interaction upon increasing accumulated electron density also suggests that the electron-electron thermalization in the gapped SS2 of the 2D TTI phase plays a significant role. Specifically, the initial relaxation dynamics of two-photon-excited electrons in the 2D TTI phase involves the electron-phonon relaxation in the $Bi_2Se_3$ bulk, as we mentioned above. However, once the relaxing electrons reach the SS2 gap, their further cooling right away prior to recombination is expected to occur through electron-electron thermalization, followed by heat generation. The smooth relaxation dynamics in the fast decay stage of the FCA response with a decay time constant of ~1.7 ps seems hence to be assigned to electron-electron thermalization and is well consistent with the metallic-like nature of Dirac SS,[16,38] although there are no gapless Dirac SS with topological protection. However, since heat transfer throughout the film is a much slower process, the carrier thermalization mechanism in the gapped SS2 of the 2D TTI phase calls for further investigations. Because the Lamb acoustic modes are possibly involved,[8] their interaction with the substrate seems to be only the way to explain this carrier relaxation dynamics.

The narrow feature appearing in the AB response of the 2D TTI phase at short delay times is due to the overlap of the negative and positive contributions, as discussed above for the 2D HTI phase [Fig. 2(b) and Fig. 7(b)]. The final relaxation dynamics in the 2D TTI phase becomes shorter compared to that in the 2D HTI phase [Fig. 7(b) and (c)]. This behavior can be associated with an increase in the gap between the Dirac cones of SS2 (as discussed in Section 3.1) and hence with the corresponding tighter suppression of the relaxation channel via the SS2 gap. The shortest decay times of the final relaxation stage (~20 ps) can hence image the actual recombination time of electrons and holes residing in the upper Dirac cone of SS2 and lower Dirac cone of SS1, respectively.

**3.4 The pumping power effect and power dependences.**

Because both the film thickness and the photoexcited carrier density play a key role on ultrafast carrier dynamics in ultrathin $Bi_2Se_3$ films (Fig. 3), the pumping power dependences of transient responses are of particular interest. As the pumping power increases, the FCA contribution to TA spectra of the 2D HTI and 2D TTI phases rises progressively with respect to the AB contribution [Fig. 8(a) and (b)], whereas the TA spectrum of 3D TI phase remains always negative and only increases in intensity [Fig. 8(c)]. This behavior additionally proves the progressive accumulation of two-photon-excited electrons in the Dirac SS2 of the 2D TTI and 2D HTI phases with increasing pumping power. The pumping power increase in the 3D TI phase also causes a redistribution of electron density from the bulk states (BS2) toward SS2, appearing as a progressive shift of the AB response peak toward ~2.0 eV [Fig. 8(c)], the energy at which the transiently excited electron population of SS2 is maximized [Fig. 6(c)]. However, the resulting electron density in SS2 of the 3D TI phase remains not high enough to appear through the inverse bremsstrahlung type FCA.

The pumping power effect on the corresponding pump-probe traces is shown in Fig. 8(d), (e), and (f). The pumping power dependences of the trace peak intensity demonstrate a nearly linear trend for the FCA response and a nearly 1/6 slope dependence for the AB response [Fig. 8(g), (h), and (i)]. These types of the pumping power dependencies image the corresponding dependencies on the photoexcited carrier density, regardless of how many photons have initially been involved within the pumping process. The pumping power dependences are hence in good agreement with the inverse bremsstrahlung type FCA and the Pauli blocking type carrier population dynamics, respectively.[25] The deviations appeared at the higher pumping powers can be associated with the progressive overlap of the negative and positive responses in the 2D TTI and 2D HTI phases and with the nonequilibrium (hot) carrier dynamics in the 3D TI phase.

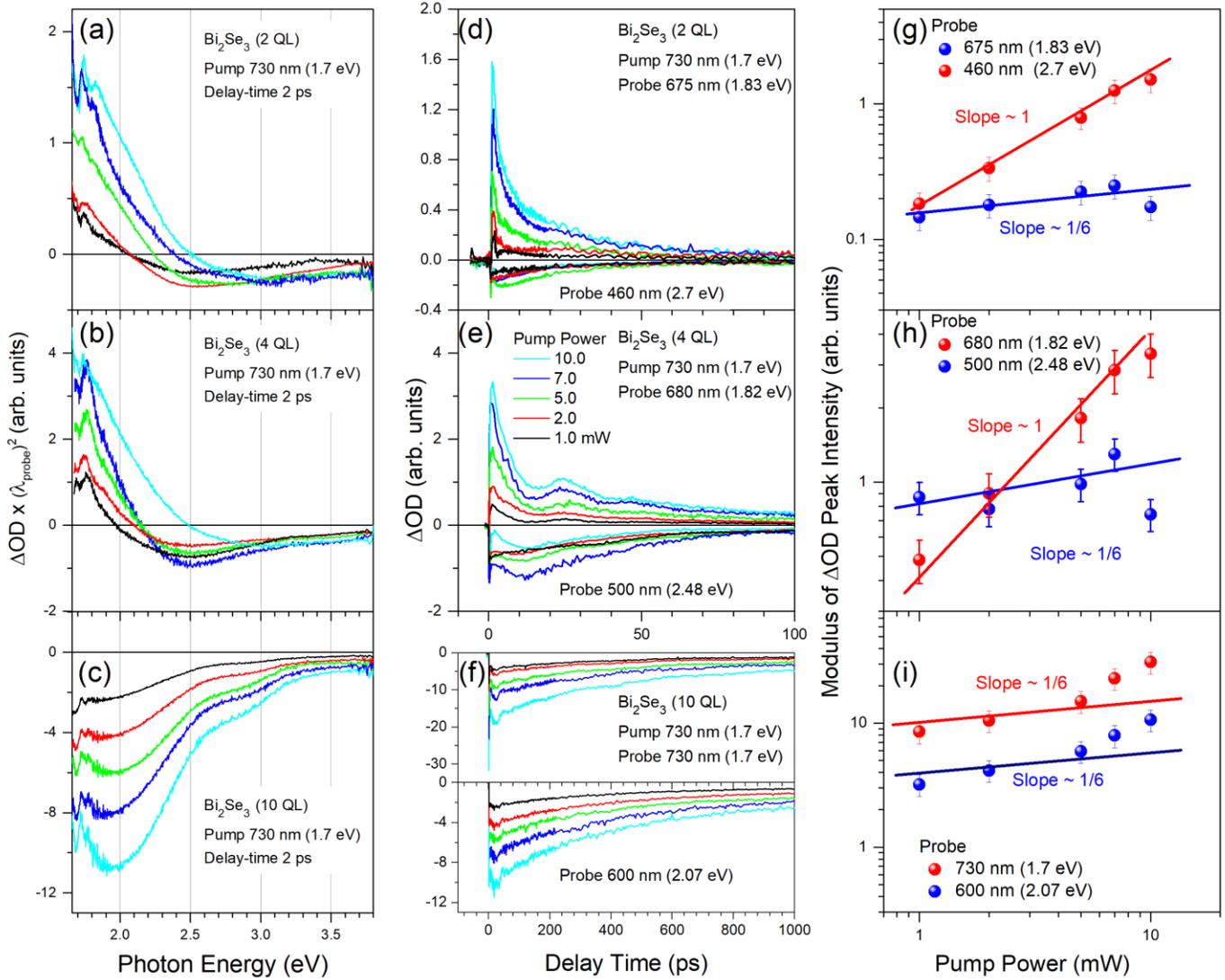

**Fig. 8.** (a), (b), and (c) A set of TA spectra for Bi₂Se₃ films with various thicknesses of 10 QL, 4 QL, and 2 QL, respectively, as a function of pumping power, as indicate by the corresponding colors. (d), (e), and (f) A set of the corresponding transient pump–probe traces. All TA spectra and traces were measured under conditions indicated for each of the panels. (g), (h), and (i) The power dependences of the modulus of the peak intensity of the transient pump–probe traces shown in (d), (e), and (f), respectively, which were plotted using log scales. The theoretically predicted slopes are shown by the corresponding color straight lines.

## 4. Conclusion

In summary, the emerging picture on the experimental findings presented in this article suggests that the ultrafast dynamics of two-photon-excited electrons in ultrathin $Bi_2Se_3$ films induced by ~1.7 eV photons is governed by the LO-phonon cascade emission in both the bulk states and Dirac SS and includes the LO-phonon-assisted SS-bulk-SS vertical transport. Because of the two-photon pumping regime with the total photon energy of ~3.4 eV, TA spectra cover the entire visible region and mainly characterize the inelastic/quasi-elastic LO-phonon-assisted relaxation of electrons prior to their reaching the Dirac point of SS2, not SS1. It is evident that to study the ultrafast carrier relaxation dynamics in SS1, either the TrARPES technique or the mid-IR/THz range TA spectroscopy is mainly required. However, as we show here, in ultrathin $Bi_2Se_3$ films with thicknesses corresponding to the 2D HTI and 2D TTI phases, the most spectacular electron relaxation dynamics occurs in the higher energy bulk states and SS2. This is a reason why the visible range TA spectroscopy becomes extremely efficient for studying ultrathin $Bi_2Se_3$ films, in view of their potential applications in novel electronic and optoelectronic devices.

Specifically, we have found that the electron relaxation dynamics is exclusively influenced by whether the Dirac point is presented between the upper and lower Dirac cones of SS2. Because in the 3D TI phase, the electron relaxation in SS2 is known to occur through two channels associated with radiative/non-radiative recombination of electrons residing in the upper Dirac cone of SS2 and holes residing in the lower Dirac cone of SS1 and with LO-phonon-assisted (or defect-assisted) scattering via the Dirac point of SS2, the Dirac SS2 act hence as a valve substantially slowing down the relaxation of electrons when the gap between the Dirac cones exceeds the LO-phonon and resonant defects energies. This behavior is exactly what is happening in the 2D HTI and 2D TTI phases. Consequently, in addition to the common AB response in TA spectra, the progressive accumulation of relaxing electrons in the upper Dirac cone of gapped SS2 leads to higher electron densities and to the positive TA response associated with the inverse bremsstrahlung type FCA.

The observed temporal evolution of the AB and FCA responses allowed us to monitor the cooling dynamics of the transiently excited electron population with time and to estimate the corresponding electron cooling and accumulation rates. As a result, we have set the electron relaxation rates in the bulk states and Dirac SS apart, which both are being caused by electron-phonon interaction, but of different type (inelastic and quasi-elastic, respectively). We found that the electron relaxation rates in the bulk states is about two times higher than that in the Dirac SS. The phonon softening effect near the surface is found to be less than 10% and therefore the softened bulk LO-phonons cannot be responsible for such a big difference experimentally observed. Consequently, we associate the observed much slower relaxation dynamics in Dirac SS compared to that in the bulk states with the much weaker electron-phonon coupling in Dirac SS due to the quasi-elastic type of electron-phonon interaction. The latter behavior is well consistent with the collisional nature of the inverse bremsstrahlung type FCA in Dirac SS. Finally, we have recognized all the specific features of the overall carrier relaxation dynamics in ultrathin $Bi_2Se_3$ films of different topological phases.

**Author contributions**

Y.D.G. modified and tested the Transient Absorption Spectrometer (TAS) (Newport), built the experimental setup, performed optical measurements, and treated the optical experimental data. J.L. assisted with the laser system operation. The optical measurements were performed in the laboratory hosted by T.H. All authors contributed to discussions. Y.D.G. analyzed the data and wrote this paper. X.W.S. guided the research and supervised the project.

**Acknowledgement**

This work was supported by the National Key Research and Development Program of China administrated by the Ministry of Science and Technology of China (Grant No. 2016YFB0401702), the Guangdong University Key Laboratory for Advanced Quantum Dot Displays and Lighting (Grant No. 2017KSYS007), the National Natural Science Foundation of China (Grant Nos. 11574130 and 61674074), the Development and Reform Commission of Shenzhen Project (Grant No. [2017]1395), the Shenzhen Peacock Team Project (Grant No. KQTD2016030111203005), and the Shenzhen Key Laboratory for Advanced Quantum Dot Displays and Lighting (Grant No. ZDSYS201707281632549). The authors acknowledge S. Babakiray for MBE growing the $Bi_2Se_3$ samples (under supervision of D. Lederman) using the West Virginia University Shared Research Facilities.